Paper ID EU-TP1852

# Neurocognitive and traffic based handover strategies

Andreas Otte[1], Jonas Vogt[1*], Jens Staub[1], Niclas Wolniak[1], Prof. Dr.-Ing. Horst Wieker[1]

1. htw saar, University of Applied Sciences, Saarbrücken, Germany, jonas.vogt@htwsaar.de

**Abstract**

The level of automation in vehicles will significantly increase over the next decade. As automation will become more and more common, vehicles will not be able to master all traffic related situations for a long time by themselves. In such situations, the driver must take over and steer the vehicle through the situation. One of the important questions is when the takeover should be performed. Many decisive factors must be considered. On the one hand, the current traffic situation including roads, traffic light and other road users, especially vulnerable road users, and on the other hand, the state of the driver must be considered. The goal is to combine neurocognitive measurement of the drivers' state and the static and dynamic traffic related data to develop an interpretation of the current situation. This situation analysis should be the basis for the determination of the best takeover point.

**Motivation**

The future of transportation is cooperative and automated driving. New concepts for the cognitive ergonomics of human-vehicle interfaces and for automated vehicles are under development. One of the basic assumptions in research and development is that people, traffic environment and vehicles increasingly form a cooperative unit for automated driving. Assistance systems make driving safer and complement the driver's natural senses. At the same time, the driver must process an ever-increasing range of information and warning messages, which results in increased distraction from the driving process [6]. The neurocognitive effects of this flood of information are the subject of current research. The neurocognitive state of the driver must be considered, when foresighted transfer strategies between the different stages of (partially) automated driving are established. Therefore, human-centered assistance systems, which evaluate the current cognitive situation of the driver, must be developed. The handover strategy from automated to manual driving and vice versa depends on the overall traffic situation. A fusion of traffic related data, sensor data and driver related data has to be performed which allows the evaluation of traffic situations regarding on their suitability for a transition of control manoeuvre.

**Approach**

A five-step approach was developed to go forward. The steps requirement analysis, data collection, data preparation, data fusion and evaluation will be described in more detail in the following sub-sections.

*Requirement Analysis*
At first, a requirement analysis is performed. It needs to be defined, what information is needed and can be gathered to capture a traffic situation as a whole. A traffic situation consists of dynamic, semi-static and static factors [1]. Static components are for example road topologies, special lanes (e.g. bus or bike lanes), intersections, or obstacles (such as buildings, billboards, gentries …) or plants that restrict visibility. Dynamic factors are other road users such as pedestrians or vehicles, current weather situation, phases of traffic lights or the neurocognitive state of the driver. Semi-static factors are construction sites or bad road conditions like potholes or road marking.

Based on the formulated requirements, scenarios were defined that describe typical situations in which a transition of the control of the vehicle manoeuvring could take place. The following four scenarios were picked out, because they contain various combinations of dynamic, semi-static and static information and therefore represent a broad variety of situations.

- Scenario 1: Stationary vehicle blocks lane
  The situation describes a broken down vehicle situation in an urban environment. The information about the stationary vehicle is provided via DSRC technology ETSI ITS-G5.
- Scenario 2: Pedestrians at intersection
  A (semi-)automated vehicle is going to turn right in a street. The intersection sensors detect multiple pedestrians, which may cross the street.
- Scenario 3: Fog
  This scenario describes a situation at a rural street with bad weather conditions including a wet and slippery street surface and dense fog.
- Scenario 4: Bad road conditions
  Due to bad road conditions, the driver in a semi-automated vehicle requests to take control of the vehicle by herself.

*Data collection*
After the definition of the requirements and scenarios, the data collection will be performed in a simulation environment as well as at the ITS test site Merzig (ITeM) [2]. The vehicle simulation software CarMaker [7] in combination with a hardware-based driving simulator is used to gather data from test drivers. The real-world test field ITeM has been recreated in this simulation environment. The real test field consists of intersections equipped with communication technology. Access to the traffic light controller allows receiving contact loop information, the current signal phases and a traffic forecast, which can be transmitted to the vehicles.

The data collection is divided in two sections: the collection of neurocognitive data and the collection of traffic-related data. The neurocognitive data is gathered by monitoring the driver. Therefore, non-contact and contact sensors are integrated into the vehicle and the driving simulator. Non-contact sensors are camera systems used for video magnification, pupillometry with eye-tracking, electrodermal response, cardiovascular parameters, and temperature recording. Contact-based sensors for electroencephalography will be used as well. The latter are not promising in their current form regarding later use in automated vehicles as the special devices could impede the driver. In a second step, after a thorough investigation which information is significant to assess the state of mind of the driver, an attempt is made to obtain the required information with a minimum set of sensors, which allow live monitoring of the driver. This is important in order to keep the cost of the system low, so that the system can also be used in series production vehicles.

The traffic-related data is gathered by infrastructure components (like road-side ITS stations and traffic lights) and vehicle components. In addition to the usual infrastructure sensors, such as contact loops, camera-based sensors are used to detect pedestrians, bicycles and other vulnerable road users (VRU). The information about the VRU is processed in the infrastructure and forwarded to other cooperative road users and to the backend. A hybrid communication approach will be deployed for this purpose. ETSI ITS-G5 technology is used for communication with other road users and infrastructure components, whereas cellular radio concepts are utilized for the connection to the backend. Additionally, in order to record test data during test drives, a pseudonymous travel time recording service will be implemented.

*Data preparation*

The characteristics of neurocognitive and traffic-related data are very different. In order to get an overall picture about the traffic situation, a fusion of those data sets is necessary. To enable a fusion of the discrete and continuous sensor data to such an overall traffic situation, the data must be unified. Therefore, a data model will be created, which enables the dimensioning and the sampling of the data. Time and position meta-information extends each data set. In traffic and communication technology, different time standards are used, which refer to different time bases and have different accuracies. For further processing of the data, this information needs to be transformed to a uniform time base. The same applies to position data. Many data from the traffic environment have no direct or global geographical reference. This reference usually results from the location of the information in field. For cloud-based processing, the spatial dimension must be added later. Not all data sources will be continuously available all the time. For these cases, an extrapolation algorithm has to be developed filling those gaps with meaningful data where necessary. All described data preparation mechanisms shall be performed on the local node where the data should be aggregated before transmitting it to the backend.

*Data fusion*

Neurocognitive and traffic based handover strategies

The data fusion is the final step in the data processing chain. It will be performed in the backend. At first, the data needs to be transferred to the backend, considering data protection and security-related aspects. Enabled by the unified data model, the fusion of the driver-related and the traffic-related data can then be performed. The aggregated and merged data can now be analysed using machine learning algorithms. The result of this analysis will be a neurocognitive normalization of a defined traffic situation allowing an automated evaluation of its suitability for the transfer of the driving task between vehicle and driver. Figure 1 visualises the three process steps: data collection, data preparation and data fusion.

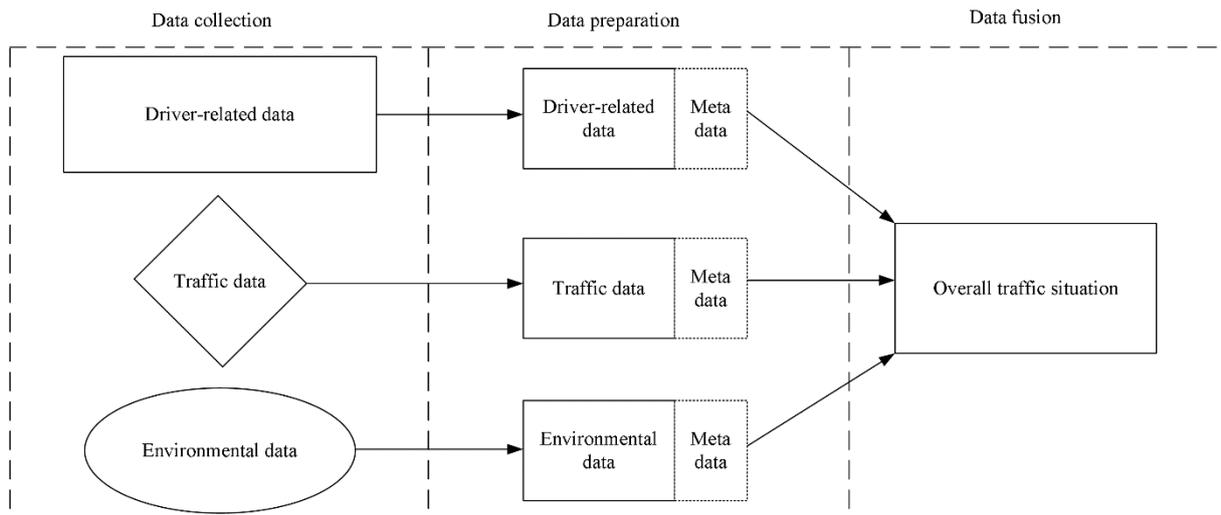

Figure 1 – Data processing chain

*Evaluation*

The results of the data fusion are evaluated in an evaluation environment. This environment is set up in the capital city of Saarland Saarbrücken with partly different conditions. There, the influence of the local conditions can be examined. With up to 140.000 external and 50.000 internal commuters, the traffic density of Saarbrücken is considerably higher than that of Merzig [3][4][5]. Saarbrücken also has a tram that creates further challenges in semi-automated driving.

**Goals**

The goal is to build up sensor infrastructure and to combine neurocognitive data with environmental and traffic-related data to evaluate the overall traffic situation. By using camera systems, which can track micro expressions in the driver's face, a detailed driver state can be detected, including emotions and vital signs. With this information, it is possible to assess traffic situations and adapt the handover strategy according to the drivers' condition. In addition, a "stress map" can be created, that indicates which areas in urban and non-urban environments are challenging for a driver of semi-automated vehicles. This information can later be provided to the vehicles and the infrastructure to make automated driving safer and more reliable.

Neurocognitive and traffic based handover strategies

**System Overview**

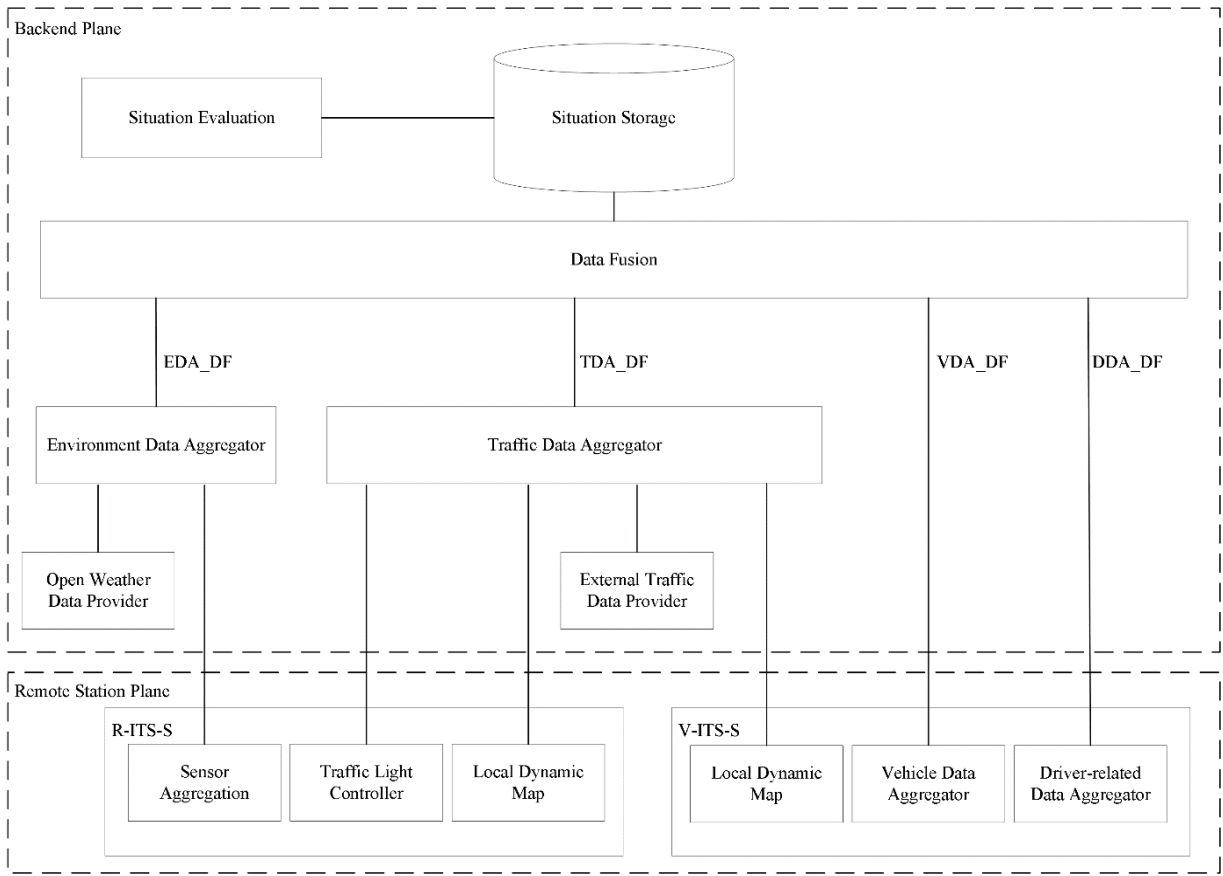

Figure 2 – System overview

Based on the approach described regarding the data collection, preparation and fusion, a detailed architecture has been created and is shown in Figure 2. The architecture consists of logical components, which are associated either with a remote station like Vehicle ITS Stations (V-ITS-S) or Road-side ITS-Stations (R-ITS-S), or the backend plane. The Local Dynamic Map (LDM) stores among others information about the traffic environment based on messages received from other traffic participants or traffic infrastructure [8]. The data of the LDM, other road events gathered by traffic control centres, and the information of the traffic light controllers of the R-ITS-S is aggregated at the Traffic Data Aggregator (TDA). The Environment Data Aggregator (EDA) collects environmental data like weather data or level of noise provided by infrastructure sensors and external data providers. Both aggregators, TDA and EDA, are services in the backend. They collect data for specific geographical areas and combine data from different sources in order to prevent data duplicates at an early stage. The data gathered by the Vehicle Data Aggregator (VDA) as well as the Driver-related Data Aggregator (DDA) cannot be combined with data from other vehicles. Therefore, there pre-aggregation is performed on the local node. The VDA has access to the data bus of the vehicle and is able to collect e.g. acceleration, speed or brake actuations. Driver-related measurements like e.g. skin conductions or eye-tracking is gathered and pre-analysed in the DDA.

The four aggregators provide the input of the Data Fusion (DF). Their task is to unify the gathered data and extend it with additional metadata. As mentioned before, the data needs a common time and location format. The chosen time format is an epoch timestamp (milliseconds since 1970-01-01) with an additional validity period. For location information, ETSI defines a format that is used to address geographical areas [9]. This format is used in the ETSI ITS-G5 messages and therefore less converting is needed, it is used here as well. The payload of the data provided by the aggregators is described as key-value-pairs. The possible keys and their corresponding data objects are defined in a data dictionary provided by the Data Fusion. Each Aggregator registers to the Data Fusion with a subset of those keys, indicating what data is provided.

The Data Fusion combines the gathered data and fills the Situation Storage (DB). The DB may be implemented as a relational database allowing links between the different data. The Situation Evaluation (EVA) can register to new situations or query the stored situations. The query mechanism allows an independent analysis of the stored situations. The results of the analysis regarding the suitability of a situation of a handover for a specific traffic participant will be stored again in the same database. The combination of the situation and their situation analysis enables further applications. For example, via an external interface, it could be possible to compare given situations with already analysed situations and to assess their suitability accordingly. The suitability of a particular intersection at a given time could also be queried.

**Conclusion**

The project will provide important insights into the transfer of control between vehicle and driver. The fusion of traffic data, vehicle sensors and neurocognitive data will allow an accurate evaluation of the overall traffic situation. Due to the use of learning machines, the transferability of the results to other situations or environments should be possible, which will be evaluated in the test field in Saarbrücken.

**Acknowledgment**

The work of this paper has been funded by the German Federal Ministry of Transport and Digital Infrastructure within the project kantSaar (grant number 16AVF2129). The project consortium consists of University of Saarland and the University of Applied Sciences Saarland. The project approach and the partial project outcomes result from the collaborative work of the entire kantSaar project team of htw saar.